\documentclass[twocolumn,showpacs,preprintnumbers,amsmath,amssymb]{revtex4}
\usepackage{graphicx}
\usepackage{dcolumn}
\usepackage{bm}
\begin{document}
\title{A simple derivation of level spacing of quasinormal frequencies for a black hole with a deficit solid angle and quintessence-like matter}


\author{Ping Xi}
\author{Xi-chen Ao}
\author{Xin-zhou Li} \email{kychz@shnu.edu.cn}
\affiliation{Shanghai United Center for Astrophysics(SUCA),
 Shanghai Normal University, 100 Guilin Road, Shanghai 200234,China}
\date{\today}
\begin{abstract}
In this paper, we investigate analytically the level space of the
imaginary part of quasinormal frequencies for a black hole with a
deficit solid angle and quintessence-like matter by the
Padmanabhan's method \cite{Padmanabhan}. Padmanabhan presented a
method to study analytically the imaginary part of quasinormal
frequencies for a class of spherically symmetric spacetimes
including Schwarzschild-de Sitter black holes which has an evenly
spaced structure. The results show that the level space of scalar
and gravitational quasinormal frequencies for this kind of black
holes only depend on the surface gravity of black-hole horizon in
the range of $-1 < w < -\frac{1}{3}$, respectively . We also extend
the range of $w$ to $w \leq -1$, the results of which are similar to
that in $-1 < w < -\frac{1}{3}$ case. Particularly, a black hole
with a deficit solid angle in accelerating universe will be a
Schwarzschild-de Sitter black hole, fixing $w = -1$ and $\epsilon^2
= 0$. And a black hole with a deficit solid angle in the
accelerating universe will be a Schwarzschild black hole,when
$\rho_0 = 0$ and $\epsilon^2 = 0$. In this paper, $w$ is the
parameter of state equation, $\epsilon^2$ is a parameter relating to
a deficit solid angle and $\rho_0$ is the density of static
spherically symmetrical quintessence-like matter at $r = 1$.
\end{abstract}


 \maketitle
\section{Introduction}
Recently, an elegant work on the quasinormal modes (QNMs) of a black
hole was done by Padmanabhan \cite{Padmanabhan}, who studied
analytically their role in response of the black hole to external
perturbation. Since the gravitational radiation excited by the black
hole oscillation is dominated by its QNMs, one can determine the
parameters of a black hole by analyzing the QNMs in its
gravitational radiation. So, besides their importance in the
analysis of the stability of the black hole, QNMs are important in
the search for black holes and their gravitational radiation. Many
physicists believe that figure of QNMs is a unique fingerprint in
directly identifying the existence of a black hole. It was found
that the structure of the spectrum of QNMs (corresponding to
quasinormal frequencies) consists of the real part and the imaginary
part \cite{Nollert,Berti} independent of initial conditions. For
example, quasinormal frequencies of Schwarzschild black hole are
\begin{equation}\label{1}
  \omega_n = i \kappa
  (n+\frac{1}{2})+\frac{\ln3}{2\pi}\kappa+O(n^{\frac{1}{2}}).
\end{equation}
where $\kappa$ is the surface gravity. The real part of quasinormal
frequencies which relates to black hole area quantization have been
studied widely \cite{Nollert,Berti,Motl}. Meanwhile, since it is
difficult to have a physical understanding of the constant spacing
of quasinormal frequencies, the imaginary part was out of
researchers' visions until recent years. Padmanabhan
\cite{Padmanabhan} presented a new analytical method (Born
approximation) which can reproduce the structure of the imaginary
part of quasinormal frequencies with evenly spacing in a general
class of spherically symmetric spacetimes for large $n$. And this
derivation can not give the real part of quasinormal frequencies.
According to the Born approximation, he came to a significant
conclusion that thermodynamics of the black hole horizon has an
effect on the imaginary part. Then, Padmanabhan and his collaborator
\cite{Choudhury}, using this method, discussed the structure of the
imaginary part of frequencies for Schwarzschild-de Sitter black
holes. They proved that although this spacetimes has two horizons,
the black hole horizon and the cosmological horizon, the imaginary
part is only related to the surface gravity of the black hole
horizon. And they explained appropriately why quasinormal modes
vanish in pure de-Sitter spacetime.

The phase transition in the early universe could have produced
different kinds of topological defects, whose cosmological
implications are very important \cite{Vilenkin,Li1,Li2}. The global
monopole, which has divergent mass in flat space-time, is one of the
most interesting defects. When one considers gravity, the linearly
divergent mass of the global monopole has an effect analogous to
that of a deficit solid angle plus a tiny mass at the origin. It has
been shown that this effective mass is actually negative
\cite{Li,Li3}. Barriola and Vilenkin point out that the metric of
global monopole with a large positive mass $M$, describes a black
hole of mass $M$ carrying a global monopole charge. Such a black
hole can be formed when a global monopole is swallowed by a
Schwarzschild black hole \cite{Barriola}. On the other hand, current
observations \cite{Ho,Hinshaw,Percival,Kowalski} (cosmic microwave
background, Type Ia Supernovae, baryon acoustic oscillation,
integrated Sachs-Wolfe effect correlations, etc.) show that there
exists a spatially homogeneous and gravitationally repulsive energy
component referred to as dark energy in our universe
\cite{Ratra,Caldwell,Hao}. One of dark energy candidates is
scalar-field dark energy models such as quintessence
($-1<w<-\frac{1}{3}$) or phantom ($w<-1$), in which $w =
\frac{p}{\rho}$ is the parameter of state equation. The solutions
have been found for a global monopole surrounded by the static
spherically-symmetric quintessence-like matter \cite{Li4,Xi}. When
such a global monopole is swallowed by an ordinary black hole, a
black hole with quintessence-like matter and a deficit solid angle
can be formed \cite{Barriola}. Therefore, it is worth further
investigating level spacing of quasinormal frequencies of this kind
of black holes analytically. Padmanabhan has pointed out that the
quantities of a physical system with a quantized spectrum are very
interesting when they have constant spacing
\cite{Padmanabhan,Choudhury}.

In this paper, through the scattering amplitude in Born
approximation \cite{Padmanabhan}, we study the level spacing of
scalar and gravitational quasinormal frequencies of in the
background of a black hole with a deficit solid angle surrounded by
quintessence-like matter, respectively. We can show that there exist
only two cases for $w < -\frac{1}{3}$. In the $w = -\frac{2k-1}{3}$
($k$ is positive integer) case, the metric function $f(r)$ has three
zero points for $-\infty < r < +\infty$. In $w \neq -\frac{2k-1}{3}$
case, $f(r)$ has only two zero points. Therefore, the spacetimes
have two horizon, the black hole horizon and the cosmological
horizon. The analytical results show that the imaginary value of
scalar and gravitational quasinormal frequencies only depend on the
surface gravity of the black hole horizon, which has an equally
spaced structure. We also consider $w \leq -1$ case, and obtain
similar results to those in $-1 < w < -\frac{1}{3}$ case.

\section{A black hole with a deficit solid angle and quintessence-like matter}

To be specific, we shall work within a particular model in unit $c =
1$, where a global $O(3)$ symmetry is broken down to $U(1)$. The
Lagrangian density is
\begin{equation}
\mathcal{L}=
\frac{1}{2}g^{\mu\nu}\partial_{\mu}\phi^{a}\partial_{\nu}\phi^{a}-\frac{\lambda^{2}}{4}(\phi^{a}\phi^{a}-{\sigma_{0}}^{2})^{2},
\end{equation}
where $\phi^{a}$ is triplet of scalar fields, and the isovector
index $a = 1, 2, 3$. The hedgehog configuration describing a global
monopole is
\begin{eqnarray}
\phi^{a} = \sigma_{0}q(\tilde{r})\frac{x^{a}}{\tilde{r}},\indent
\textrm{with}\indent x^{a}x^{a}=\tilde{r}^{2}.
\end{eqnarray}
so that we shall actually have a monopole solution if $q \rightarrow
1$ at spatial infinity and $q \rightarrow 0$ near the origin. The
solutions for a global monopole surrounded by the static
spherically-symmetric quintessence-like matter are as follows
\cite{Li4}:
\begin{eqnarray}
ds^2&=&(1-\frac{2G\sigma_0m}{r}-\epsilon^2+\frac{\rho_{0}}{{3w}}r^{-3w-1})dt^2\nonumber\\
&-&\frac{1}{1-\frac{2G\sigma_0m}{r}-\epsilon^2+\frac{\rho_{0}}{{3w}}r^{-3w-1}}dr^2\nonumber\\
&-&r^2(d\theta^2+\sin^2\theta d\phi^2),
\end{eqnarray}
where $\epsilon\equiv\sqrt{8\pi G\sigma_0^2}$ is a dimensionless
parameter of a deficit solid angle, $w < -\frac{1}{3}$,
$m\approx-\frac{16\pi\sigma_0}{3\lambda}$ and $\rho_0$ is the
density of static spherically symmetrical quintessence-like matter
at $r = 1$.

When such a global monopole is swallowed by an ordinary black hole
with mass \~{M}, a black hole with a deficit solid angle surrounded
by quintessence-like matter can be formed:
\begin{equation}
ds^2 = f(r)dt^2-\frac{1}{f(r)}dr^2 -r^2d\Omega^2,
\end{equation}
where $f(r) =
(1-\frac{2M}{r}-\epsilon^2+\frac{\rho_{0}}{{3w}}r^{-3w-1})$,
$M=G\sigma_0$(\textit{\~{M}}$-m$) is the dimensionless parameter of
mass of a black hole with a deficit solid angle surrounded by
quintessence-like matter. The necessary condition is
\begin{equation}
\rho_{0} <
(1-\epsilon^2)[\frac{(1-\epsilon^2)(3|w|-1)}{6M|w|}]^{3|w|-1}.
\end{equation}
for the existence of a black hole. For example, we consider $w =
-\frac{1}{2}$ case. If $\rho_{0} <
(1-\epsilon^2)\sqrt{\frac{1-\epsilon^2}{6M}}$, there exists the
solution of a black hole with a deficit solid angle surrounded by
quintessence-like matter. And if $\rho_{0} \geq
 (1-\epsilon^2)\sqrt{\frac{1-\epsilon^2}{6M}}$, there exists the
 naked singularity solution.

The function $y(r)$ is given by
\begin{eqnarray}\label{3}
\frac{y(r)}{r} &=& f(r)\nonumber\\
&=&1-\frac{2M}{r}-\epsilon^2+\frac{\rho_0}{3w}r^{-3w-1},
\end{eqnarray}
Next, we discuss an equation as follows
\begin{equation}\label{1}
y(r) = 0,
\end{equation}
Obviously, the roots of $f(r) = 0$ are identical to those of Eq.
(\ref{1}). Then, we will discuss the real roots of $y(r) = 0$. In
the range of $0 < r < +\infty$, there is only one stationary point
at $\tilde{r} = (\frac{1-\epsilon^2}{\rho_0})^{-\frac{1}{3w+1}}$.
From the necessary condition (6), we have $y(\tilde{r}) > 0$. Since
$y''(\tilde{r})=(3w+1)\rho_0 < 0$ (the prime denotes the derivative
with respect to $r$.) and $y(0) = -2M$, we conclude that the
equation $y(r) = 0$ has and only has two positive real roots. That
means the spacetime has two horizons, the black hole horizon
($r=r_b$) and the cosmological horizon ($r = r_c$, where $r_b <
r_c$). Next, we will consider whether there exist negative real
roots of $y(r) = 0$. If $r$ is real and negative, we have
\begin{eqnarray}\label{5}
r^{-3w} &=& (-1)^{-3w}|r|^{-3w}\nonumber\\
 &=& [\cos{(-3w)\pi}+i\sin{(-3w)\pi}]|r|^{-3w},
\end{eqnarray}
Therefore, Eq. (\ref{1}) can be reduced to
\begin{equation}
(-1)^{-3w-1}\frac{\rho_0}{3w}|r|^{-3w}+(1-\epsilon^2)|r|+2M = 0.
\end{equation}
Using a similar method, we find that there is a negative root iff $w
= -\frac{2k-1}{3}$. But in $w \neq -\frac{2k-1}{3}$ case, the real
roots are not negative for Eq. (\ref{1}).

\section{Quasinormal modes}
\subsection{The $-1 < w < -\frac{1}{3}$ Case}

Now, we consider concretely the behaviors of scalar perturbations in
a black hole with quintessence-like matter and a deficit solid
angle. The propagation of a massless scalar field is described by
the Klein-Gordon equation
\begin{equation}\label{6}
\nabla_{\mu}\nabla^{\mu}\Phi = 0\indent (\mu = 0,1,2,3).
\end{equation}
Then we separate variables by setting
\begin{equation}\label{7}
\Phi(t,r,\theta, \phi) = \frac{1}{r}\psi(r)Y_{lm}(\theta,
\phi)e^{i\omega t},
\end{equation}
where $Y_{lm}(\theta, \phi)$ are the usual spherical harmonics.
Submitting Eqs. (\ref{7}) to (\ref{6}), we obtain
\begin{equation} \label{8}
\frac{d^2\psi(r)}{dr_\ast^2}+(\omega^2-V_s)\psi(r) = 0,
\end{equation}
where $r_\ast$ is the tortoise coordinate
\begin{eqnarray}\label{9}
r_\ast &\equiv& \int{\frac{1}{f(r)}}dr\nonumber\\
&=&
\frac{1}{2\kappa_b}\ln{|\frac{r}{r_b}-1|}-\frac{1}{2\kappa_c}\ln{|1-\frac{r}{r_c}|},
\end{eqnarray}
and $V_s$ is the effective potential
\begin{equation}\label{10}
V_s = f(r)
[\frac{l(l+1)}{r^2}+\frac{2M}{r^3}+\frac{\rho_{0}(3w+1)}{3w}r^{-3w-3}],
\end{equation}
where $\kappa_b$ and $\kappa_c$ represent the surface gravity of the
black hole horizon and the surface gravity of cosmological horizon,
respectively.

For gravitational perturbations, the metric function is expressed as
\begin{equation}
g_{\mu\nu}=\bar{g}_{\mu\nu}+h_{\mu\nu},
\end{equation}
where $\bar{g}_{\mu\nu}$ is the background metric, and $h_{\mu\nu}$
is a small perturbation. Here, We adopt the canonical form for
$h_{\mu\nu}$ in classical Regge-Wheeler gauge \cite {Regge} {\small{
\begin{displaymath} h_{\mu\nu}= \left(\begin{array}{cccc}
0&0&0&h_{0}(r)\\
0&0&0&h_{1}(r)\\
0&0&0&0\\
h_0(r)&h_1(r)&0&h_0(r)
\end{array}\right)e^{-i\omega t}(\sin\theta\frac{\partial}{\partial\theta})P_{l}(\cos\theta),
\end{displaymath}}

Introducing
$Q(r)=\frac{1-\frac{2M}{r}-\epsilon^2+\frac{\rho_0}{3w}r^{-3w-1}}{r}h_1(r)$,
we obtain
\begin{equation}
\frac{d^2Q(r)}{dr_\ast^2}+(\omega^2-V_g)Q(r)=0,
\end{equation}
where $V_g$ is the effective potential
\begin{equation}
V_g=
f(r)[\frac{l(l+1)}{r^2}-\frac{6M}{r^3}+\frac{\rho_{0}(3w+1)}{3w}r^{-3w-3}],
\end{equation}

For scalar and gravitational perturbations, we can show that the
effective potentials $V_s$ and $V_g$ vanish at two horizons, which
correspond to $r_\ast \rightarrow -\infty$ ($r = r_b$) and $r_\ast
\rightarrow +\infty$ ($r = r_c$). So, the wave functions as
solutions for Eq. (13) and Eq. (17) can be plane wave as follows
\begin{eqnarray}\label{11}
\psi \sim \left\{\begin{array}{ll}
 e^{i\omega r_\ast}&\textrm{$r_\ast \rightarrow
-\infty$},\\
e^{-i\omega r_\ast}&\textrm{$r_\ast \rightarrow +\infty$}.
\end{array}\right.
\end{eqnarray}
and
\begin{eqnarray}\label{31}
Q \sim \left\{\begin{array}{ll}
 e^{i\omega r_\ast}&\textrm{$r_\ast \rightarrow
-\infty$},\\
e^{-i\omega r_\ast}&\textrm{$r_\ast \rightarrow +\infty$}.
\end{array}\right.
\end{eqnarray}
where $\omega$ denotes quasinormal frequencies, the imaginary part
of which are identical to the poles of the scattering amplitude
$S(\omega)$ in momentum space.

The scattering amplitude in the Born approximation, as the Fourier
transform of potential $V({\bf x})$ in momentum space, is
\begin{equation}\label{12}
S({\bf q}) = \int{d{\bf x}V({\bf x})e^{-i{\bf qx}}},
\end{equation}
where ${\bf q} = {\bf k_f}-{\bf k_i}$ is the momentum transfer. In
one dimension, we take ${\bf k_f} = -{\bf k_i}$, then ${\bf q} =
-2{\bf k_i}$. Thus, the scattering amplitude can be written as
\begin{equation}\label{13}
S(\omega) = \int^{+\infty}_{-\infty}{dr_\ast V(r(r_\ast))e^{i2\omega
r_\ast}},
\end{equation}
Submitting Eqs. (\ref{9})-(\ref{10}) to Eq. (\ref{13}), we get
\begin{eqnarray}\label{14}
S(\omega) &=& \int^{r_c}_{r_b} dr
[\frac{l(l+1)}{r^2}+\frac{2M}{r^3}+\frac{\rho_0(3w+1)}{3w}r^{-3w-3}]\nonumber\\
&\times& (\frac{r}{r_b}-1)^{i\frac{\omega}{\kappa_b}}(1-\frac{r}{r_c})^{-i\frac{\omega}{\kappa_c}}\nonumber\\
&=& 2MI_3+l(l+1)I_2+\frac{\rho_0(3w+1)}{3w}I_{3w+3},
\end{eqnarray}
Submitting Eq. (14) and Eq. (18) to Eq. (22), we get
\begin{eqnarray}\label{14}
S(\omega) &=& \int^{r_c}_{r_b} dr
[\frac{l(l+1)}{r^2}-\frac{6M}{r^3}+\frac{\rho_0(3w+1)}{3w}r^{-3w-3}]\nonumber\\
&\times& (\frac{r}{r_b}-1)^{i\frac{\omega}{\kappa_b}}(1-\frac{r}{r_c})^{-i\frac{\omega}{\kappa_c}}\nonumber\\
&=& -6MI_3+l(l+1)I_2+\frac{\rho_0(3w+1)}{3w}I_{3w+3},
\end{eqnarray}
where
\begin{eqnarray}\label{15}
I_N &=& \int^{r_c}_{r_b}{dr r^{-N}
(\frac{r}{r_b}-1)^{i\frac{\omega}{\kappa_b}}(1-\frac{r}{r_c})^{-i\frac{\omega}{\kappa_c}}}\nonumber\\
&=&
(\frac{r_c-r_b}{r_b})^{i\frac{\omega}{\kappa_b}}(\frac{r_c-r_b}{r_c})^{-i\frac{\omega}{\kappa_c}}r_{b}^{-N}\nonumber\\
&\times&
F_1(1+i\frac{\omega}{\kappa_b},N,2+i\omega[\frac{1}{\kappa_b}-\frac{1}{\kappa_c}];-\frac{r_c-r_b}{r_b})\nonumber\\
&\times& \frac{\Gamma
{(1+i\frac{\omega}{\kappa_b})}\Gamma{(1-i\frac{\omega}{\kappa_c})}}{\Gamma{(2+i\omega[\frac{1}{\kappa_b}-\frac{1}{\kappa_c}])}}~~
(N=3w+3, 2, 3),
\end{eqnarray}
In the expression of $I_N$, the ratio of
$\frac{F_1}{\Gamma{(2+i\omega[\frac{1}{\kappa_b}-\frac{1}{\kappa_c}])}}$
doesn't have poles anywhere \cite{Gradshteyn}. Although the factor
$\Gamma{(1-i\frac{\omega}{\kappa_c})}$ has poles at $Im(\omega_n) =
-n\kappa_c (n \gg 1)$, we know that $-in\kappa_c$ is not the
imaginary value of quasinormal frequencies from the definition of
quasinormal modes. Therefore, we infer that the pole structure of
scattering amplitude is given by
\begin{equation}\label{16}
S(\omega) \propto \Gamma{(1+i\frac{\omega}{\kappa_b})},
\end{equation}
It means that the imaginary part of quasinormal frequencies for
scalar and gravitational perturbations is given by
\begin{equation}\label{17}
Im(\omega_n) = n\kappa_b ~~~~~~(n \gg 1).
\end{equation}
It is clear that the level spacing of the imaginary parts of scalar
and gravitational quasinormal frequencies are determined by the
surface gravity at black hole horizon in $-1 < w < -\frac{1}{3}$
case, and it is equally spaced.

We have used the first Born approximation to obtain the QNMs
Spectrum for the black hole with a deficit solid angle and
quintessence-like matter. The approximation gives the correct level
spacing for the imaginary values of the QN frequencies. As we know,
QN frequencies include the real and imaginary parts. However, the
analytical expression of the real part is difficult to obtain. The
numerical studies of the real part can be carried out by WKB-like
approximation techniques. By numerical calculations, we attain that
$\frac{Im(\omega_{n+1})}{Im(\omega_n)} \approx 1+\frac{1}{n}$ ($n
\gg 1$) for scalar and gravitational perturbations in the $-1 < w <
-\frac{1}{3}$ case showed in Tables 1-2, which is consistent with
the result of the formula (27). Next, we will discuss QNMs of this
kind of black holes in the range of $w \leq -1$.

\begin{table*}
\caption{Scalar QN frequencies of a black hole with a deficit solid
angle and quintessence-like matter for $M=1$, $\epsilon^2=0.001$,
$\rho_0=0.01$, $l=3$ and $w = -\frac{2}{3}$.}
\begin{tabular}{cc}
 \hline
 $n$ &$\omega_n$ \\ \hline
 3&0.58889-0.67804i
 \\
4&0.55178-0.88128i
 \\
5&0.50894-1.08641i
\\
6&0.45997-1.29294i
\\
7&0.40455-1.50083i
\\
8&0.34249-1.71015i
\\
9&0.27373-1.92109i
\\
10&0.19828-2.13380i
\\
\hline
\end{tabular}
\end{table*}
\begin{table*}
\caption{Gravitational QN frequencies of a black hole with a deficit
solid angle and quintessence-like matter for $M=1$,
$\epsilon^2=0.001$, $\rho_0=0.01$, $l=3$ and $w = -\frac{2}{3}$.}
\begin{tabular}{cc}
 \hline
 $n$ &$\omega_n$ \\ \hline
 3&0.50553-0.65551i
 \\
4&0.46253-0.85301i
 \\
5&0.41235-1.05270i
\\
6&0.35457-1.25427i
\\
7&0.28895-1.45781i
\\
8&0.21542-1.66350i
\\
9&0.13402-1.87158i
\\
10&0.04486-2.08225i
\\
\hline
\end{tabular}
\end{table*}

\subsection{The $w \leq -1$ Case}
In the $w \leq -1$ case, there are two subcases. From the analysis
of the real roots of $f(r) = 0$, we know that there are only two
positive roots in $w \neq -\frac{2k+1}{3}$ subcase, similar to that
in the range of $-1 < w < -\frac{1}{3}$. In $w = -\frac{2k+1}{3}$
subcase, there are two positive
 and one negative real roots of $f(r) = 0$. The tortoise coordinate
 $r_\ast$ is
{\small\begin{equation}\label{18} r_\ast =
\frac{1}{2\kappa_b}\ln{|\frac{r}{r_b}-1|}-\frac{1}{2\kappa_c}\ln{|1-\frac{r}{r_c}|}+\frac{1}{f'(r_h)}\ln{|1+\frac{r}{|r_h|}|},
\end{equation}}
where $r_h$ is the negative root, which satisfies
\begin{equation}\label{19}
\frac{\rho_0}{3w}(r_{c}^{2k+1}+r_{b}^{2k+1}-2r_{n}^{2k+1})+(1-\epsilon^2)(r_c+r_b-2r_h)=0,
\end{equation}\label{19}
According to Abel theorem, we can not find the exact form by the
finite algebra operation, except for $k = 1$. For $k = 1$, we obtain
the relation as follows
\begin{equation}\label{20}
r_h = -(r_c+r_b).
\end{equation}
which is similar to that in Schwarzschild-de Sitter spacetime
\cite{Choudhury}. By the way, if $\rho_0 = 0$ (there is no
quintessence) and $\epsilon^2 = 0$, $f(r)$ is the same as the metric
of Schwarzschild black hole. Submitting Eqs. (\ref{10}) and
(\ref{18}) to Eq. (\ref{13}), we get
\begin{eqnarray}\label{21}
S(\omega) &=& \int^{r_c}_{r_b} dr
[\frac{l(l+1)}{r^2}+\frac{2M}{r^3}+\frac{\rho_0(3w+1)}{3w}r^{-3w-3}]\nonumber\\
&\times& (\frac{r}{r_b}-1)^{i\frac{\omega}{\kappa_b}}(1-\frac{r}{r_c})^{-i\frac{\omega}{\kappa_c}}(1+\frac{r}{|r_h|})^{i\frac{\omega}{f'(r_h)}}\nonumber\\
&=& 2MI_3+l(l+1)I_2+\frac{\rho_0(3w+1)}{3w}I_{3w+3},
\end{eqnarray}
Submitting Eqs. (18) and (28) to Eq. (\ref{13}), we get
\begin{eqnarray}\label{21}
S(\omega) &=& \int^{r_c}_{r_b} dr
[\frac{l(l+1)}{r^2}-\frac{6M}{r^3}+\frac{\rho_0(3w+1)}{3w}r^{-3w-3}]\nonumber\\
&\times& (\frac{r}{r_b}-1)^{i\frac{\omega}{\kappa_b}}(1-\frac{r}{r_c})^{-i\frac{\omega}{\kappa_c}}(1+\frac{r}{|r_h|})^{i\frac{\omega}{f'(r_h)}}\nonumber\\
&=& -6MI_3+l(l+1)I_2+\frac{\rho_0(3w+1)}{3w}I_{3w+3},
\end{eqnarray}
where {\small \begin{eqnarray}\label{22} I_N &=& \int^{r_c}_{r_b}{dr
r^{-N}
(\frac{r}{r_b}-1)^{i\frac{\omega}{\kappa_b}}(1-\frac{r}{r_c})^{-i\frac{\omega}{\kappa_c}}(1+\frac{r}{|r_h|})^{i\frac{\omega}{f'(r_h)}}}\nonumber\\
&=&
(\frac{r_c-r_b}{r_b})^{i\frac{\omega}{\kappa_b}}(\frac{r_c-r_b}{r_c})^{-i\frac{\omega}{\kappa_c}}(\frac{r_b+|r_h|}{|r_h|})^{i\frac{\omega}{f'(r_h)}}r_{b}^{-N}\nonumber\\
&\times&
F_1(1+i\frac{\omega}{\kappa_b},N,i\frac{\omega}{f'(r_h)},2+i\omega[\frac{1}{\kappa_b}-\frac{1}{\kappa_c}];-\frac{r_c-r_b}{r_b}\nonumber\\&,&-\frac{r_c-r_b}{r_b+|r_h|})\nonumber\\
&\times& \frac{\Gamma
{(1+i\frac{\omega}{\kappa_b})}\Gamma{(1-i\frac{\omega}{\kappa_c})}}{\Gamma{(2+i\omega[\frac{1}{\kappa_b}-\frac{1}{\kappa_c}])}}~~~
(N=3w+3, 2, 3),
\end{eqnarray}}
The ratio of
$\frac{F_1}{\Gamma{(2+i\omega[\frac{1}{\kappa_b}-\frac{1}{\kappa_c}])}}$
also doesn't have poles anywhere \cite{Gradshteyn}. It is easy to
find that the poles of scattering amplitude are only determined by
the poles of the function $\Gamma{(1+i\frac{\omega}{\kappa_b})}$.
Thus, the imaginary part of scalar and gravitational quainormal
frequencies in $w = -\frac{2k+1}{3}$ case are also given by
\begin{equation}\label{23}
Im(\omega_n) = n\kappa_b ~~~~~~(n \gg 1).
\end{equation}
The level spacing of the imaginary part of frequencies for scalar
and gravitational perturbations only depends on the surface gravity
of the black hole horizon in $w = -\frac{2k+1}{3}$ case, which is
supported by our numerical results listed in Tables 3-4.
\begin{table*}
\caption{Scalar QN frequencies of a black hole with a deficit solid
angle and quintessence-like matter for $M=1$, $\epsilon^2=0.001$,
$\rho_0=0.01$, $l=3$ and $w = -\frac{4}{3}$.}
\begin{tabular}{cc}
 \hline
 $n$ &$\omega_n$ \\ \hline
 3&0.56082-0.63141i
 \\
4&0.52960-0.81378i
 \\
5&0.49091-0.99737i
\\
6&0.44485-1.18239i
\\
7&0.39153-1.36903i
\\
8&0.33109-1.55749i
\\
9&0.26368-1.74794i
\\
10&0.18946-1.94053i
\\
\hline
\end{tabular}
\end{table*}
\begin{table*}
\caption{Gravitational QN frequencies of a black hole with a deficit
solid angle and quintessence-like matter for $M=1$,
$\epsilon^2=0.001$, $\rho_0=0.01$, $l=3$ and $w = -\frac{4}{3}$.}
\begin{tabular}{cc}
 \hline
 $n$ &$\omega_n$ \\ \hline
 3&0.48784-0.61217i
 \\
4&0.45249-0.78844i
 \\
5&0.40824-0.96629i
\\
6&0.35535-1.14609i
\\
7&0.29408-1.32814i
\\
8&0.22470-1.51269i
\\
9&0.14750-1.69993i
\\
10&0.06263-1.89005i
\\
\hline
\end{tabular}
\end{table*}
\section{Conclusion}
It is known that QNMs' frequencies of black holes including
Schwarzschild and Schwarzschild-de Sitter are equally spaced, with
the level spacing depending only on the surface gravity. In this
paper, we generalize this result to a new class of spacetimes and
provide the imaginary parts of scalar and gravitational quasinormal
frequencies by Pabmanabhan's method. In the view of extensibility
and simplicity, Padmanabhan's analysis is an elegant method.
Furthermore, these analyses show that the result is closely related
to the thermal nature of horizons and is the consequence of the
exponential redshift of the wave modes close the horizon. In $w =
-\frac{2k+1}{3}$ case, there are three real roots of $f(r) = 0$, one
of which is negative. In $w \neq -\frac{2k+1}{3}$ case, there are
two positive real roots. However, the imaginary part of quasinormal
frequencies of a black hole with a deficit solid angle for scalar
and gravitational perturbations in these two cases are only
determined by the surface gravity at black-hole horizon, which are
equally spaced. In particular, a black hole with a deficit solid
angle in an accelerating universe will be a Schwarzschild-de Sitter
black hole, fixing $w = -1$ and $\epsilon^2 = 0$. And a black hole
with a deficit solid angle in accelerating universe will be a
Schwarzschild black hole,when $\rho_0 = 0$ and $\epsilon^2 = 0$.



\begin{acknowledgments}
We thank Prof. T. Padmanabhan for the helpful suggestion to study
the level spacing of quasinormal frequencies for black holes with a
deficit solid angle surrounded by quintessence-like matter. This
work is supported by National Science Foundation of China grant No.
10847153.
\end{acknowledgments}

\end{document}